\documentclass[twocolumn,showpacs,preprintnumbers,amsmath,amssymb]
{revtex4}

\usepackage{epsfig}

\newcommand{\be}{\begin{equation}}
\newcommand{\ee}{\end{equation}}

\begin{document}

\title{Granular clustering self-consistent analysis for general
coefficients of restitution}

\author{E. Thiesen and W.A.M. Morgado}\altaffiliation{Corresponding
 author: W.A.M.Morgado\\ welles@fis.puc-rio.br}
  \affiliation{Departamento de
 F\'{\i}sica, Pontif\'{\i}cia Universidade Cat\'olica do Rio de
 Janeiro \\ CP 38071, 22452-970 Rio de Janeiro, Brazil}

\date{\today}

\begin{abstract}
We study the equilibrium behavior of one-dimensional granular
clusters and one-particle granular gases for a variety of velocity
dependent coefficients of restitution $r$. We obtain equations
describing of the long time behavior for the cluster's pressure,
r.m.s. velocity and granular inter-spacing. We show that for
extremely long times, clusters with velocity dependent
coefficients of restitution are unstable and dissolve into
homogeneous, quasi-elastic gases, but clusters with velocity
independent $r$ are permanent.  This is in accordance with
hydrodynamic studies pointing to the transient nature of density
instabilities for granular gases with velocity dependent $r$.
\end{abstract}

\maketitle

\newline


\section{Introduction}

Granular materials are present in many natural systems and play an
important role in our daily lives~\cite{PT1,duran}, and in the
economy since an estimated 1.3\% of the U.S. electric power
consumption goes into grinding particles and ores~\cite{enge1}.
The interest in such systems ranges from the purely theoretical to
daily practical applications, such as in construction
industry~\cite{duran}.

Typical granular systems (GS) are composed of large numbers of
discrete macroscopic grains. Their shape is usually irregular but
in the present work we will consider them as smooth regular
spheres of diameter $d$ in vacuum, as a good approximation that
still captures some of the essential physics of the problem.
However, their dynamical and statistical properties may be
affected by the presence of an interstitial medium, such as air or
a liquid when their Bagnold number is small
enough~\cite{mobius,duran}. Granular materials behave in
interesting ways exhibiting different features from ordinary
solids, liquids and gases, such as arches redistributing loads to
the sides of solid arrangements of grains and inelasticity-induced
non trivial velocity distributions in rapid flowing granular
gases. Even some basic laws of thermodynamics, such as the zeroth
law, may fail when extended to GS~\cite{feitosa,breyT,brey1}.

A very important aspect of the behavior of GS is their inherent
tendency to cluster, i.e., a compaction due to an enhancement of
the rate of collisions inside the system, accompanied by granular
cooling down (kinetic energy reduction). Many authors have studied
granular gas clustering, or compaction for denser systems,
behavior for GS which are initially homogeneous and start with a
given amount of kinetic
energy~\cite{GHZ,brey1,ben-naim,tsimring,nie1}, and
subsequently left on their own to cool down. Their initial cooling
behavior, the homogeneous cooling state (HCS), obeys Haff's
Law~\cite{haff,nie1} for the granular temperature (the typical
internal average kinetic energy for the grains).

For longer times and rarefied granular systems, the inter-granular
collisions tend to correlate the motion (velocities and positions)
of the grains and techniques from Kinetic Theory will not be
reliable anymore, at least in its simplest
form~\cite{tan,precorr,kada2}. Theoretical models have developed upon
this notion and obtained scaling forms for the transport
coefficients~\cite{ertas1,ertas2} when the velocities of masses of
grains become correlated inside a region of a certain
characteristic length.

For clusters coalesced from smooth granular gases (no tangential
restitution) there are no mechanisms for the exchange of angular
momentum and rolling does not occur (a totally irrelevant sliding
motion between grains may still be present, but the rotational
motion of the grains is not coupled to the translational motion if
the grains are smooth). It is legitimate, from a theoretical point
of view, to ask whether such structures are really permanently
stable or some other mechanism could lead the system again to
gaseous homogeneity.

It has been known, for quite some time, that the hydrodynamic
approximation predicts density instabilities for inelastic, smooth,
hard-sphere granular systems at zero gravity~\cite{brey4} (velocity
independent coefficients of restitution) while for equivalent systems
with velocity dependent coefficient of restitution ($r$), such as the
viscoelastic model~\cite{viscoelastic} the instabilities are only
transient~\cite{tpnbsw}. Thus, the dependence of the coefficient of
restitution on the velocity might be the cause of a possible cluster
break-up.  That is due to the fact that $r$ tends to 1 as the impact
relative velocity tends to zero.  However, to simulate a cluster
break-up can become computationally very costly, if no approximations
are used. In most event driven Molecular Dynamics simulations the
coefficient of restitution has to be set to 1 as the relative velocity
becomes smaller than an elastic threshold, so that collisions at
relative velocities lower than the threshold will be
elastic~\cite{tpnbsw,nie1}.  In order to be more realistic, one has to
be able to reach extremely long times and small velocities. This is
the goal of the present work. We will study a simple qualitative model
(a 1-dimensional cluster) and obtain the asymptotic solutions for its
behavior at extremely long times, not yet accessible to computer
simulations.

We study systems with coefficients of restitution given by the general
form~\cite{edson}, at small relative velocity $g$
\be
r=1-A\left(\frac{g}{g_0}\right)^{m},
\label{coefrest}
\ee where $0\leq A <1$, and the constant $g_0$ sets the dissipation
scale, and $m\geq 0$. The cases $m=0$ (velocity independent $r$) and
$m=1/5$ (viscoelastic)~\cite{viscoelastic} are particular cases of
Eq.~\ref{coefrest}.

We assume that the system starts as a granular gas inside a bounded
1-dimensional container by walls at the extremities: one elastic and
one inelastic. A cluster forms at the inelastic end and a gas still
exists at the elastic side after some initial large time
interval~\cite{kada1,kada2}. The granular gas portion will consist
ultimately of a single grain. We will use this assumption in what
follows but this is not really essential, and our results could easily
be extended to multi-grain gases, provided the gas kinetic energy goes
asymptotically to zero.

We will obtain equations of motion for the dynamic and
quasi-thermodynamic granular quantities: gas velocity, the cluster
granular temperature (in fact its r.m.s. velocity) and internal
pressure. We will show that the cluster is unstable at extremely
long times, except for the case of velocity-independent case
($m=0$).  Purely dynamic effects, due to energy dissipation, can
be responsible for cluster formation but can not hold it together
indefinitely, except for the unrealistic velocity-independent
coefficient of restitution case.

This paper is organized as follow. In Section II, we write the basic
equations of collision dynamics. In Section III, we present our model
and its corresponding equations. In Section IV, we study the stability
conditions for the cluster. In Section V, the long time behavior is
extensively analyzed. In Sections VI and VII, we discuss some of the
consequences of the model and make our concluding remarks. The more
technical aspects of this work are to be found in Appendices
~\ref{apa} and ~\ref{apb}.

%
%
\section{Basic Assumptions}
Our approach is qualitative in nature and we do not expect it to
reproduce the detailed behavior of a true granular cluster but only a
few of its crudest properties.

Our model consists of an unforced one dimensional system with $N+1$
identical smooth grains, with unit mass $M=1$, which can be
modeled by inelastic hard-spheres with velocity dependent
coefficient of restitution $r$, given by Eq.~\ref{coefrest}.

\begin{figure}[h]
\includegraphics[scale = 0.3]{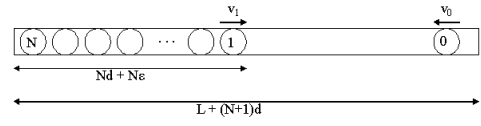}
\caption{The $N$ particles of the cluster and the gas particle.
The average distance between two consecutive particles of the
cluster is $\varepsilon$ (not shown), the diameter of each
particle is $d$, and the total length is $L+(N+1)d$.}\label{fig1}
\end{figure}

The grains are confined in a region of length $L+(N+1)d$ ($d$ is the
granular diameter) by an elastic wall on the right and an inelastic
one on the left, as shown in figure~\ref{fig1}.  Relative to each
other, the elastic wall presents some similarities to a ``hot'' wall
while the inelastic one would be a ``cold'' one, as in
reference~\cite{cordero}, since the inelastic one is where the energy
is taken out of the system. However, we are not injecting any amount
of energy into the system and a steady-state does not develop.  The
inelastic collisions at the inelastic wall are governed by the same
Eq.~\ref{coefrest}, the wall being an infinitely heavy grain.  The
velocities of two colliding grains will be $(V_1',V_2')$ after the
collision, and $(V_1,V_2)$ before it. They are related by: \be
V_{1}'=\left(\frac{1-r}{2}\right)V_{1}+\left(\frac{1+r}{2}\right)V_{2},
\label{v1} \ee \be
V_{2}'=\left(\frac{1+r}{2}\right)V_{1}+\left(\frac{1-r}{2}\right)V_{2}.
\label{v2} \ee

The system starts with a given initial amount of kinetic energy that
will be dissipated due to the internal collisions. A partial
clustering of grains will initially occur at the inelastic wall due to
the pressure of the remaining gas~\cite{kada1}. The cluster phase is
formed when the relative velocities are large compared with
$g_0$ and the coefficient of restitution differs appreciably from 1. The gas will lose
particles to the cluster until only one gas particle is
left~\cite{kada2}. The difference between
clustering and collapse is important, since in the later an infinite number of
collisions occur among the particles in a finite amount of
time. However, the cluster is only a very dense concentration of grains
with small relative speeds. In one dimension clustering can precede
collapse (for $r$ close to 1)~\cite{sela}.

After some initial interval of time, the velocities of all grains,
cluster's one and the gas one, will be much smaller than the inelastic
velocity scale $g_0$ in Eq.~\ref{coefrest}.

We label the gas particle as the $0^{th}$ particle while the cluster
ones are labeled from 1 to N. We assume $g_0\gg v_{0}\gg
v_{i=1,\cdots, N}$.  Although the velocity of the gas particle is much
larger than the velocities of the particles forming the cluster, the
scaling factor $g_0$ will also be a lot larger than the velocity of
the gas particle for long times.  The following hierarchy for the
small expansion parameters holds (for $m>0$):
\begin{equation}
\left|\frac{v_{i}}{v_{0}}\right|\,\gg\,
\left|\frac{v_{0}}{g_0}\right|^{m}\,\gg\,
\left(\frac{v_{i}}{v_{0}}\right)^{2}\,\gg\,
\left(\frac{v_{i}}{v_{0}}\right) \left|\frac{v_{0}}{g_0}\right|^{m}.
\label{init}
\end{equation}

The logic of Eq.~\ref{init} is that at sufficient long times
$\frac{v_{i}}{g_0}\rightarrow 0$ and (since the gas will keep pumping
energy into the cluster) the ratio $v_i/v_0$ will not vanish (that
will be checked a posteriori, with respect to the long time behavior
of the cluster). We can assume (for $m > 0$) that
$\left|v_{i}/v_{0}\right|\,\gg\, \left|v_{0}/g_0 \right|^{m}$. The
second inequality comes from a choice that the time we choose to start
our calculations is large but not too large, since for even larger
times, certainly $\left|v_{i}/v_{0}\right|^2\,\gg\, \left|v_{0}/g_0
\right|^{m}$. The last inequality is a direct consequence of this
choice of initial time.  Higher order terms shall be discarded.

We assume that our model
has initial conditions satisfying Eq.~\ref{init}.

%
%
\section{Gas-cluster equilibrium}
\subsection{Gas pressure}
In order to calculate the pressure exerted by the gas on the
cluster we need to take into account the momentum exchanged
between the gas particle and the cluster after each collision
between gas and cluster. A gas-cluster collision is completed when
the gas particle leaves the cluster with an outgoing velocity
(which is close in absolute value to its velocity before
collision) much larger than the typical cluster's particle's
velocity. This is illustrated in Fig.~\ref{fig1}.  We notice that
as the gas particle collides with the cluster, it is as if the
``fast particle'' (gas) would pass through the cluster's ones,
collide with the inelastic wall and go all the way back in the
direction of the elastic wall.  The process can be repeated until
the ``fast particle'' crosses the whole cluster, back and forth.
Thus, we obtain, to leading order, the momentum exchanged between
the gas and cluster. This is calculated in Appendix~\ref{apa},
Eq.~\ref{vdot} (notice that $V$ is the speed of the incoming gas
particle, $V\geq0$ always):
\begin{eqnarray}
\dot{V} &=&-(2N+1)\frac{A}{4L}\left|\frac{V}{g_0}\right|^{m}V^2,
\label{eq1}
\end{eqnarray}
where the flipping of the ``fast particle's'' velocity at the
inelastic wall is correctly taken into account in the final result
above.

From Eq.~\ref{pgas} ($m>0$), we obtain
\begin{equation}
p_{gas}=\frac{\Delta P_{cluster}}{\Delta t} = \frac{(1+m)}{2}
\frac{\dot{V}}{V}\,\,\dot\varepsilon,\label{eq22}
\end{equation}
where $p_{gas}$ is the pressure exerted by the gas on the cluster, and
$\Delta P_{cluster}$ is the inelastic change of cluster momentum due
to the collision with the gas particle.

For $m=0$, Eq.~\ref{pgas0} gives us
\begin{equation}
p_{gas}=\frac{1-(1-A)^N}{1+(1-A)^{N}}\,\, \dot V.\label{eq222}
\end{equation}

We will assume in the following that the product $NA$ is small. Our results,
being valid for small $NA$ in the $m=0$ case, will show that collapse happens
in a finite amount of time. This will be certainly valid for the case of large $NA$.

\subsection{Cluster's variables}
Other important variables exist that we need to take into account.
The variance of the cluster's particle's velocities $\sigma^{2}$ is
one of them. It is defined as
\begin{equation}
\sigma^2=\frac{\sum_{i=1}^{N}(v_i-v_{cm})^2}{N},\label{sigma1}
\end{equation}
where the cluster's center of mass velocity is given by
\begin{equation}
v_{cm}=\frac{\sum_{i=1}^{N}v_i}{N}.
\end{equation}

Another such variable is the mean granular spacing, $\varepsilon$. It
is defined by (with $x_{N+1}=0$)
\begin{equation}
\varepsilon=\frac{\sum_{i=1}^{N+1}\left|x_i-x_{i-1}-d\right|}{N}.
\label{varepsilon1}
\end{equation}
The cluster's full size is thus $N(d+\varepsilon)$. A somewhat crude, but reasonable,
approximation for the cluster's center of mass velocity can be
obtained by assuming that the cluster expands or contracts rather
uniformly (at least for small $\varepsilon$), and thus
\begin{eqnarray}
v_{cm}&\approx &\sum_{i}^{N}Ni\dot{\varepsilon}=
\frac{N(N+1)}{2N}\dot{\varepsilon}.\label{vcm1}
\end{eqnarray}
In the limit of large $N$ the expression above reduces to
\[
v_{cm}\approx\frac{N}{2}\dot{\varepsilon}.
\]

The center of mass' acceleration then reads
\begin{equation}
a_{cm}=\frac{N(N+1)}{2N}\ddot{\varepsilon}.\label{eq2}
\end{equation}

\subsection{Wall pressure and mean spacing}
The cluster feels two external sources of pressure: the pressure
from the gas and the pressure due to interactions with the
inelastic wall. That pressure can be calculated using a crude
approximation by assuming that the momenta exchanged between the
particle labeled $N$ at every collision is of the order of
$2\sigma$ and the rate of collision is
$\sigma/\varepsilon$~\cite{chap}. The mean-field wall pressure is
then
\begin{equation}
p_{wall} = \frac{2\sigma^2}{\varepsilon}>0.\label{eq3}
\end{equation}

The equation of time evolution for the mean spacing $\varepsilon$
is obtained by writing Newton's law for the cluster
($p_{wall}$ is a positive force while $p_{gas}$ is a negative one):
\[
a_{cm}=p_{wall}+p_{gas}.
\]
Using Eqs.~\ref{eq22},~\ref{eq2} and ~\ref{eq3} we obtain the
mean-field equation for $m>0$:
\begin{equation}
\frac{N}{2}\ddot{\varepsilon} =
\frac{2\sigma^2}{\varepsilon}+\frac{(1+m)}{2}
\frac{\dot{V}}{V}\,\,\dot\varepsilon.
\label{eq5}
\end{equation}

The equivalent equation for the case $m=0$ is obtained from
$p_{gas}$ given by Eq.~\ref{pgas0}
\begin{equation}
\frac{N}{2}\ddot{\varepsilon} =
\frac{2\sigma^2}{\varepsilon}+\frac{1-(1-A)^N}{1+(1-A)^{N}}\,\,\dot
V. \label{eq50}
\end{equation}

\subsection{Energy dissipation inside the cluster}
A calculation similar to that for the gas velocity reduction is done
in Appendix~\ref{apb} for the energy of the cluster, due to the effect
of gas-cluster collision. At the order of approximation we have set,
$|V/g_0|^m$, the corrections will be of the order
$\sigma^2|V/g_0|^{2m}$ (much smaller than $\sigma^2/\varepsilon$).
The comparison of the cluster's kinetic energy before and after a
collision with the gas particle then reads
\begin{equation}
\sum_{i}^{N}(v_{i}'''')^{2}=\sum_{i}^{N}(v_{i})^{2}.
\end{equation}
The cluster's kinetic energy is not affected by the gas-cluster
collision, at the order of approximation used. This tells us that only
internal collisions will be important for cooling down the cluster.

In a mean field approximation, the energy dissipated per particle
corresponds to the product of the energy lost in each collision and
the rate of collision per particle. For a $N$-particle one-dimensional
gas confined in a free volume $l_{1D}$, with the typical velocity
variance $\sigma$ ($v$ is the typical velocity, of the order of the
square root of the variance $\sigma$), the energy loss per collision
corresponds roughly, in dimensionless terms, to
\[
\Delta v \propto -\left|v\right|^{1+m}.
\]
The rate of collision is proportional to $v/l_{1D}$~\cite{chap}.  So,
in order to calculate the rate of dissipation of energy inside the
cluster, with typical inter-spacing becoming very small as
$t\rightarrow\infty$, but not yet zero, we need to estimate the
internal collision rate and multiply it by the loss of energy for each
collision. The rate of collisions for a cluster only differs from that
of a gas because its inter-spacing $\varepsilon$ is very small
compared to $L$. Taking it into account, the cluster's collision rate
is now obtained as
\[
q=N\frac{\sigma}{2\varepsilon}.
\]
The variation of the cluster's typical internal velocity is due to
the collision between grains. We assume that colliding
grains will have relative velocities of the order $\sigma$ and the
quasi-elastic collisions (in the limit $g\rightarrow0$) will
switch those velocities with a small loss
\[
\Delta \sigma = - A\left(\frac{\sigma}{g_0}\right)^m\sigma.
\]
The change in $\sigma$ per unit time is the product
$q\Delta\sigma$. Thus we obtain the heuristic equation:
\begin{equation}
\dot{\sigma} =
-\frac{NA\sigma^2}{2\varepsilon}\left(\frac{\sigma}{g_0}\right)^m.
\label{eq4}
\end{equation}

The equivalent form for $m=0$ is derived from Eq.~\ref{dotsigma} from
appendix B:
\begin{equation}
\dot{\sigma} \sim
-\frac{\sigma^2}{\varepsilon}.
\label{eq40}
\end{equation}

%
%
\section{Dimensionless analysis}
We want to study, qualitatively, the conditions for the cluster to
be stable and shall use the approximate equations of motion
obtained above. However, we will not be interested in the fine
details of the equations themselves, only in their asymptotic
behavior in time. Then, we will rewrite
Eqs.~\ref{eq1},~\ref{eq5},~\ref{eq50} and~\ref{eq4} in a
completely dimensionless form as (notice that $V,
\sigma,\varepsilon > 0$ for all $t$). First, the equations for $V$
and $\sigma$ have the same form for both $m=0$ and $m>0$. They
read
\begin{eqnarray}
\dot{V} &=&-V^{2+m},
\label{e1}\\
\dot{\sigma} &=& -\frac{\sigma^{2+m}}{\varepsilon}. \label{e2}
\end{eqnarray}
The equation for $\ddot\varepsilon$ for $m=0$ reads
\begin{eqnarray}
\ddot{\varepsilon} &=&
\frac{\sigma^2}{\varepsilon}+\dot{V},\label{e30}
\end{eqnarray}
and that for $m>0$ reads
\begin{eqnarray}
\ddot{\varepsilon} &=&
\frac{\sigma^2}{\varepsilon}+\frac{\dot{V}}{V}\,\,
\dot\varepsilon.\label{e3}
\end{eqnarray}

In order to recover the dimensional units, remember that $\sigma$ and $V$
are given in terms of $g_0$, and $\varepsilon$ is measured in terms of $L/N$.
A few dimensional constants have to be used in Eqs.~\ref{e1}-\ref{e3} in
order to make both sides dimensionally coherent.

Eq.~\ref{e1} can be exactly solved, obtaining
\begin{equation}
V =
\frac{V_0}{\left(1+(1+m)V_0^{1+m}t\right)^{\frac{1}{1+m}}}.\label{e5}
\end{equation}
This is the extension of Haff's law~\cite{haff,poeschel33} to the cases
described by Eq.~\ref{coefrest}. We observe that as
\[
t\rightarrow\infty \Rightarrow V\sim t^{-\frac{1}{1+m}} \Rightarrow
T_g\sim t^{-\frac{2}{1+m}}.
\]
For the case of velocity independent coefficient of restitution,
$m=0$, $T_g\sim t^{-2}$. For the viscoelastic coefficient of
restitution case~\cite{viscoelastic}, $m=1/5$, $T_g\sim
t^{-\frac{5}{3}}$ as expected~\cite{poeschel33}.  Eqs.~\ref{e2},
~\ref{e3} and~\ref{e30} will constitute the system to be solved in
the following.

%
%
\section{Long-time behavior}
We must keep in mind that there is an implicit velocity scale
$g_0$ that divides the velocities variables whenever a power of
$m$ comes into play (a consequence of the form of the coefficient
of restitution). For the constant coefficient of restitution case,
we obtain a useful equation by combining Eq.~\ref{e2} with
Eq.~\ref{e30}:
\begin{equation}
\dot\varepsilon-V+\frac{\sigma^{1-m}}{1-m}=c_0.\label{ccc}
\end{equation}
where the constant $c_0$ is related to the initial conditions for
$V$, $\sigma$ and $\varepsilon$ by:
\begin{equation}
c_0=\dot\varepsilon_{0}-V_{0}+\frac{\sigma_0^{1-m}}{1-m}.\label{c0}
\end{equation}
The equations describing the granular cluster are valid in the
limit when $V\gg \sigma,\dot\varepsilon$.  In the following we
analyze the cluster behavior for the whole range of values of $m$
based on Eqs.~\ref{e1}-\ref{ccc}. The figures are obtained by
solving numerically the Eqs.~\ref{e1}-\ref{ccc}.

\subsection{$m=0$}
In this case, the coefficient of restitution does not depend on the
impact relative velocity. We have:
\begin{eqnarray}
c_0  &=& \dot\varepsilon_0-V_{0}+\sigma_{0}\approx -V_{0}\\
\dot\varepsilon &=& V-\sigma-V_{0}.
\end{eqnarray}
As $t\to\infty$, we observe that $V$ and $\sigma$ tend to
zero~\cite{footnote1} and thus $
\dot\varepsilon\approx-V_{0} < 0$.

In practice, it is impossible to observe the asymptotic limit above
since the collapse happens on a finite amount of time (see
figure~\ref{migual0}). The cluster is thus stable and will not
dissolve itself.

\subsection{$m>0$}
When the coefficient of restitution depends on the initial relative
velocity, i.e. $m>0$, we can see that the physical behavior of the
system changes qualitatively, as shown in Appendix~\ref{apa}.

A scaling argument can be used and compared with the result of
simulations in order to obtain the very long time behavior of the
variables $V$, $\sigma$ and $\varepsilon$. The asymptotic solutions
for $\varepsilon$ and $\sigma$ can be written as powers of time and log-time:
\begin{equation}
V\sim t^{\beta_{1}}, \hspace{1cm}\sigma\sim t^{\beta_{2}}(\ln
t)^{\alpha_2},
\hspace{1cm}\varepsilon\sim t^{\beta_{3}}(\ln
t)^{\alpha_3},\label{ansatz}
\end{equation}
where we have already determined $\beta_1$:
\[
\beta_1=-\frac{1}{1+m}.
\]

The solutions for $\alpha_2$, $\beta_2$, $\alpha_3$ and $\beta_3$ are
\begin{eqnarray}
\alpha_2  &=& -\frac{1}{m},\\
\beta_{2} &=&0, \\
\alpha_3  &=& -\frac{1}{m},\\
\beta_{3} &=& 1.
\end{eqnarray}

Thus, the long time behavior of $\sigma$ and $\varepsilon$ is given by
\begin{eqnarray}
\sigma\sim (\ln t)^{-\frac{1}{m}},\,\,\mbox{ and }
\varepsilon\sim t(\ln t)^{-\frac{1}{m}}.\label{assintotas}
\end{eqnarray}

These are self-consistent, logarithmically corrected solutions for
$\sigma$ and $\varepsilon$ at long times.  However, we need to look
into the long-times behavior of $\sigma$ with more detail.

For $m>0$ there is no granular collapse. After some transient time,
the cluster will grow almost linearly (as can be seen in
figure~\ref{fig_todas}, consistent with the main behavior of
$\varepsilon\sim t$) and will eventually occupy the whole container,
in fact becoming once again a granular gas with interspacing
$\varepsilon\sim L/N$. It takes an enormous amount of time for this to
happen.  This is illustrated in figure~\ref{fig_todas} where we
compare the cases $m=0.2$, $m=1$ and $m=2$.

At long-times, our model becomes quasi-elastic, to a very good
approximation, when $m>0$. For the velocity dependent case, the
internal dissipation for the cluster becomes negligible
($\beta_2=0$) but its internal energy is not conserved. The
apparent contradiction between $\beta_2=0$ and $\sigma\rightarrow
0$ does not hold since $\sigma$ decays as a power of $\ln t$. Even
more significantly, in our model $\varepsilon \leq L/N$ and the
growth of $\varepsilon$ has to be cut-off correspondingly. Since
our mean-field equations do not impose a boundary to
$\varepsilon$, then Eq.~\ref{e2} will give us
$\dot\sigma\rightarrow 0$ as $t\rightarrow \infty$, consistent
with $\beta_2=0$. In reality, after reaching the cut-off size,
normal gas dissipation takes over and the former cluster will
follow Haff's law for energy dissipation again.

Another important consistency argument can be extracted from
Eq.~\ref{assintotas}. If we take the limit $m\rightarrow0$ before the
limit $t\rightarrow\infty$ is taken, we observe that
$\varepsilon=\sigma=0$ results. This is in complete agreement (remember that the
initial time is taken to be  long for the $m>0$ case) with
our result that a granular collapse happens in a finite interval of
time when $m=0$.

It is interesting to notice at this point that in
reference~\cite{tpnbsw} the authors simulate a velocity dependent
granular system with an elastic threshold ($r=1$ below a certain
threshold relative velocity) and supposed their results to be
extensible to the viscoelastic regime. This is in
accordance with our results. However, the form of the coefficient of
restitution in reference~\cite{tpnbsw} much more closely mimics the
$m>1$ than the case $m=0.2$. As we observe in our calculations,
$\varepsilon$ will neither tend to zero nor remain stable in both
cases, which is consistent with the results in reference~\cite{tpnbsw}.

\subsection{Oscillations}
An interesting feature we have observed are very low frequency
size-oscillations, at very long times. Eq.~\ref{e2} predicts
oscillations with decreasing frequency. We can study the case of a
small perturbation on $\varepsilon$ such as:
\[
\varepsilon=t (\ln t)^{-\frac{1}{m}}\left[1+\phi\right].
\]

We obtain, after some straightforward algebra, the asymptotic
equation for the relative perturbation $\phi$:
\begin{equation}
\ddot{\phi}+\frac{\ln t}{t}\dot\phi+\frac{2\ln t}{t^2}\phi
=0.\label{HO}
\end{equation}
It is similar to a low frequency damped harmonic oscillator, with
a frequency that goes to zero as $t^{-1}(\ln t)^{\frac12}$.

The effect of small perturbations in the asymptotic value of
$\varepsilon$ is rather hard to observe directly. However, we
observed it by initially running our simulations in order to
obtain aged values of $\varepsilon$, $\sigma$ and its derivatives.
We then perturb $\varepsilon$ as $(1+\Delta)\varepsilon$, with
$\Delta = 1.0\times10^{-4}$.

We run two subsequent calculations, with an aged and unperturbed
solution as the initial condition, and another for the perturbed
one. Their difference should also obey Eq.~\ref{HO}. The result is
plotted in figure~\ref{amplia}, in a logarithmic scale (we plot the
absolute value of $\phi$; the signs correspond to whether $\phi$ is
positive or negative).

It can be seen that the period is indeed increasing (it is of the
order of the total time, consistent with a ``frequency'' of order
$t^{-1}(\ln t)^{\frac12}$.

%
%
\section{Consequences of the model}
The most immediate consequence of the present model (for $m\neq0$)
is the evidence it provides of the transient nature for some of
the granular singularities in a freely cooling granular gas with
velocity-dependent coefficient of restitution. This indicates that
a hydrodynamic treatment might be adequate for such systems, at
least after a transient time. Also, we deduce from our results
that purely dynamical effects cannot give rise to permanent
clusters if $m\neq0$ at the zero-energy feeding regime.

Another consequence is the eventual evaporation of clusters for smooth
granular systems. The inviscid Burgers' equation has been proposed as
a mechanism of formation for a granular cluster with velocity
independent coefficient of restitution~\cite{doolen,nie1}. For systems
with $m\neq0$ one may ask whether that equation is still adequate, and
what kind of regime might replace it, in the evaporative period (at
extremely long times). Work is currently under way along this
direction.

The non-collapse when $m\neq0$ gives us hope that it might be
possible to treat two- or three-dimensional clusters as a very
dense, but non-singular, granular phases (for smooth systems)
describable by internal, non-diverging, variables (maybe even
similar ones to the $\sigma$ and $\varepsilon$ used in this
manuscript). That could make it easier to incorporate the
treatment of clusters into the hydrodynamic methods available
today.

%
%
\section{Conclusions}
We study the long term stability of unforced granular systems, in
which clusters form, with the help of a qualitative, microscopic
model that makes it possible to look at clusters at extremely long
times, not available to computer simulations.

We assume a general form for the coefficient of restitution that
includes the well known velocity-independent and viscoelastic models
as special cases.

We are interested in this problem for two main reasons. Firstly,
despite its apparent simplicity, a granular cluster's behavior, at
extremely long times, depends on the amount of inelasticity (which can
be defined as $q=\frac{1-r}{2}$). According to our model, if the
coefficient of restitution becomes 1 as the relative velocity of
impact tends to zero, as with most realistic systems, then clusters of
rigid, smooth spheres will be unstable (at least at zero
gravity). This suggests a rich dynamical behavior for our granular gas
that comprises an initial homogeneous phase in which Haff's
law~\cite{haff} predicts the evolution of the average granular
temperature. The system goes into phase separation after a transient
time and the global kinetic energy varies with a different power of
time~\cite{nie1}. After a very long waiting time, the external
granular gas pressure no longer keeps the cluster particles together
and the cluster finally dissolves into an extremely slow moving
homogeneous granular gas. Haff's law will once again apply to this gas
(since $m\neq0)$. This is not in contradiction with the results in
reference~\cite{doolen} since the results therein apply to systems
with velocity independent coefficients of restitution ($m=0$).

Secondly, for velocity dependent coefficients of restitution, the
clusters are not truly collapsed, but behave instead as very
dense, fluid phases (for zero surface friction and zero gravity).
In fact, we could think of the gas-cluster phase coexistence
boundary as a smooth separation between the granular gas and
cluster phases, without a singular boundary, except for the case
of constant coefficients of restitution ($m=0$). An appropriate
continuous hydrodynamic treatment for it might be possible.

Questions arise concerning the long times dissolution of granular
clusters: will they obey the same equations as the ones that are
found to apply for the collapsing phase? Since the irreversibility
of the ``microscopic'', e.g. granular, dynamics prevents
time-reversal to apply, the dissolution equations might be quite
different from the collapse ones. This is yet to be understood.


%
%
\section{Gas-cluster momentum exchange}\label{apa}
\subsection{$m>0$}
From Eqs.~\ref{v1} and~\ref{v2}, we obtain the post-collision
velocities for two particles of the same mass (the collision time
is taken to be zero):
\begin{equation}
v_{1}'=\left(1-\,\frac{A}{2}
\left|\frac{v_{1}-v_{0}}{g_0}\right|^{m}\right) v_{0}
+\frac{A}{2}\left|\frac{v_{1}-v_{0}}{g_0}\right|^{m}v_{1},
\label{v1-1}
\end{equation}
\begin{equation}
v_{0}''=\frac{A}{2}\left|\frac{v_{1}-v_{0}}{g_0}\right|^{m}v_{0}+
\left(1-\,\frac{A}{2}\left|\frac{v_{1}-v_{0}}{g_0}\right|^{m}\right)
v_{1}. \label{v2-1}
\end{equation}

The final velocity of the ``fast particle'' after crossing the
cluster can be calculated by the equations of basic collision
dynamics. Assuming $g_0\gg |v_{0}|=V\gg |v_{1}|$, we can expand
the last term of Eq.~\ref{coefrest} as follows:
\begin{eqnarray}
\nonumber \left|\frac{v_{1}-v_{0}}{g_0}\right|^{m} &=&
        \left|\frac{v_{0}}{g_0}\right|^{m}\left|1-\,
\frac{v_{1}}{v_{0}}\right|^{m}\\
        &\cong& \left|\frac{V}{g_0}\right|^{m}
        \left(1+m\,\frac{v_{1}}{V}\right).
\end{eqnarray}

The velocities of particle 1 and the gas particle, after the first
collision, can be rewritten with the help of Eq.~\ref{v1-1} and
Eq.~\ref{v2-1} as:
\[
v_{1}' \approx
          -V+\frac{A}{2}\left|\frac{V}{g_0}\right|^{m}V
          +(1+m)\frac{A}{2}\left|\frac{V}{g_0}\right|^{m}v_{1},
\]
now being the fast particle, and the gas one is now slow (primes
stand for fast-slow collisions):
\[ v_{0}'' \approx
          v_{1}-\frac{A}{2}\left|\frac{V}{g_0}\right|^{m}V
          -(1+m)\frac{A}{2}\left|\frac{V}{g_0}\right|^{m}v_{1}.
\]

After $\ell$ collisions, the fast particle velocity will be the
$\ell$-th one
\[
v_{\ell}' \approx
          -V+\ell\,\frac{A}{2}\left|\frac{V}{g_0}\right|^{m}V
          +(1+m)\frac{A}{2}\left|\frac{V}{g_0}\right|^{m}
          \sum_{i=1}^{\ell-1}v_{i},
\]
and the $(\ell-1)$-th particle (which suffered two collisions)
has the velocity:
\[
v_{\ell-1}'' \approx
          v_{\ell}-\frac{A}{2}\left|\frac{V}{g_0}\right|^{m}V
          -(1+m)\frac{A}{2}\left|\frac{V}{g_0}\right|^{m}v_{\ell}.
\]

After colliding N times, the fast particle will reach the
inelastic wall. Its velocity, prior to that collision, will read
then:
\begin{eqnarray}
\nonumber v_{N}'
      &=& -V+N\frac{A}{2}\left|\frac{V}{g_0}\right|^{m}V+
               (1+m)\frac{A}{2}\left|\frac{V}{g_0}\right|^{m}p_{cl},
\end{eqnarray}
where $p_{cl}=\sum_{i=1}^{N}v_{i}$, the cluster's total momentum
before colliding with the gas.

Hence the total momentum given by the gas to the cluster (first
part)
\begin{eqnarray}
\nonumber \Delta p_{cluster_1}
      &=& -N\frac{A}{2}\left|\frac{V}{g_0}\right|^{m}V
               -(1+m)\frac{A}{2}\left|\frac{V}{g_0}\right|^{m}p_{cl}.
\end{eqnarray}

After the collision with the inelastic wall, it reads:
\begin{eqnarray}
\nonumber V' &=& v_{N}''\\ \nonumber
      &=& V-(N+1)\frac{A}{2}\left|\frac{V}{g_0}\right|^{m}V-
               (1+m)\frac{A}{2}\left|\frac{V}{g_0}\right|^{m}p_{cl}.
\end{eqnarray}
The procedure for the calculations of how the fast particle
traverses the cluster is similar to the one above and the final
result is (up to the same approximation order)
\begin{eqnarray}
\nonumber v_{0}'''
      &=& V'-N\frac{A}{2}\left|\frac{V}{g_0}\right|^{m}V'+
               (1+m)\frac{A}{2}\left|\frac{V}{g_0}\right|^{m}p_{cl}'',
\end{eqnarray}
where $p_{cl}''=\sum_{i=1}^{N}v_{i}''$.

The momentum received by the cluster due to these collisions is
then
\begin{eqnarray}
\nonumber \Delta p_{cluster_2}
      &=& N\frac{A}{2}\left|\frac{V}{g_0}\right|^{m}V'-
               (1+m)\frac{A}{2}\left|\frac{V}{g_0}\right|^{m}p_{cl}'',\\
               \nonumber
      && \hspace{-1.5cm}=N\frac{A}{2}\left|\frac{V}{g_0}\right|^{m}V-
      (1+m)\frac{A}{2}\left|\frac{V}{g_0}\right|^{m}p_{cl}'',
\end{eqnarray}

Before the collision with the elastic wall, the fast particle
velocity is given by:
\begin{eqnarray}
\nonumber v_{0}''' &=&
      V'-N\frac{A}{2}\left|\frac{V}{g_0}\right|^{m}V'+
      (1+m)\frac{A}{2}\left|\frac{V}{g_0}\right|^{m}p_{cl}'',\\
      \nonumber &=&
      V-(2N+1)\frac{A}{2}\left|\frac{V}{g_0}\right|^{m}V+\\ \nonumber
      &&+(1+m)\frac{A}{2}\left|\frac{V}{g_0}\right|^{m}(p_{cl}''
      -p_{cl}),
\end{eqnarray}

After that last collision, the gas particle has the velocity:
\begin{eqnarray}
\nonumber v_{0}'''' &=&
      -V+(2N+1)\frac{A}{2}\left|\frac{V}{g_0}\right|^{m}V-\\ \nonumber
      &&-(1+m)\frac{A}{2}\left|\frac{V}{g_0}\right|^{m}(p_{cl}''-p_{cl}),
\end{eqnarray}
Hence, the absolute value of the gas particle velocity varies (for a
single gas-cluster collision cycle) as:
\begin{eqnarray}
\Delta V &=&
-(2N+1)\frac{A}{2}\left|\frac{V}{g_0}\right|^{m}V\nonumber \\ &&
+(1+m)\frac{A}{2}\left|\frac{V}{g_0}\right|^{m}(p_{cl}''-p_{cl}),
\nonumber\\ &=&
-(2N+1)\frac{A}{2}\left|\frac{V}{g_0}\right|^{m}V,\label{ce1}
\end{eqnarray}
where we discarded terms of the order ${\cal
O}(\left|V/g_0\right|^{2m})$.

Thus, the total momentum absorbed by the cluster from the gas-cluster
collision is given by
\begin{eqnarray}
\nonumber \Delta p_{cl\,tot}&=&\Delta p_{cluster_1}+\Delta
      p_{cluster_2}\\ &=&
      -(1+m)\frac{A}{2}\left|\frac{V}{g_0}\right|^{m}
      (p_{cl}''+p_{cl})\nonumber \\ &=&
      -(1+m)A\left|\frac{V}{g_0}\right|^{m}p_{cl}.\label{ce2}
\end{eqnarray}
The reader should notice that we assume the initial time to be
large enough so that quantities such as $NA|V/g_{0}|^{m}$ are
small and the total dissipation per gas-cluster collision can be a
small fraction of the gas kinetic energy.

Equations~\ref{ce1} and ~\ref{ce2} are the fundamental result of
this appendix. We can transform them into rate equations by
determining the rate of gas-cluster collisions. The time interval
between successive collisions is given by $\Delta t=2L/V$.

We obtain the equation governing the behavior of the absolute
value of the gas velocity:
\begin{eqnarray}
\dot V \equiv \frac{\Delta V}{\Delta t} &=&
      -(2N+1)\frac{A}{4L}\left|\frac{V}{g_0}\right|^{m}V^2. \label{vdot}
\end{eqnarray}

The equation for the gas pressure, the rate of transfer of
momentum is also obtained
\begin{eqnarray*}
p_{gas}\equiv \frac{\Delta p_{cl\,tot}}{\Delta t}
      &=& -(1+m)\frac{A}{2L}\left|\frac{V}{g_0}\right|^{m}Vp_{cl}.
\end{eqnarray*}
Notice that for a large cluster, $N\gg1$, we can write
\begin{eqnarray}
p_{gas}
      &=&\frac{(1+m)}{2}\frac{\dot
      V}{V}\,\,\dot\varepsilon \,.\label{pgas}
\end{eqnarray}

\subsection{$m=0$}\label{apa2}
The case of a constant coefficient of restitution deserves a
separate  treatment. In this case we assume that $A\ll 1$ and
$r=1-A\approx 1$.

After a collision with a slow particle, the fast particle acquires
a velocity $v' = (1-A)v.$ At the end of a sequence of N such
collisions, the velocity of the fast particle (before colliding
with the inelastic wall) will be
\[
v_N=(1-A)^Nv_0.
\]
The momentum exchanged with the cluster is then
\[
\Delta p_{c1}=\left((1-A)^N-1\right)v_0.
\]

After colliding with the inelastic wall, the gas particle has a
velocity $v_N=-(1-A)^Nv_0$. After colliding another N times with
cluster particles, the gas particle velocity will be
\[
v_N''=-(1-A)^{2N}v_0.
\]
The momentum exchanged with the cluster this time is then
\[
\Delta p_{c2}=-(1-A)^{N}\left((1-A)^N-1\right)v_0.
\]

The total change in velocity for the gas particle, after collision
with the elastic wall, is given by
\begin{equation}
\Delta V = -\left(1-(1-A)^{2N}\right)V,
\end{equation}
The rate of change of $V$ is given (see the calculation for $m>0$
above)
\begin{equation}
\dot V = -\left(\frac{1-(1-A)^{2N}}{2L}\right)V^2.\label{vdot0}
\end{equation}

The total momentum gained by the cluster after the collision is
then:
\begin{equation}
\Delta p_{c} = \Delta p_{c1}+\Delta p_{c2}=
-\left((1-A)^N-1\right)^2 V.
\end{equation}
The gas pressure in this case will be given by (similarly to case
for $m>0$)
\begin{equation}
p_{gas}=\frac{\Delta p_{c}}{\Delta t} =
-\frac{\left((1-A)^N-1\right)^2}{2L} V^2.\label{pgas00}
\end{equation}

We can see that the gas pressure is related to the change in gas
velocity through
\begin{equation}
p_{gas} = \frac{1-(1-A)^N}{1+(1-A)^{N}}\,\, \dot V. \label{pgas0}
\end{equation}

There is a clear change in the gas pressure regime for $m>0$ compared
with the more commonly used case of $m=0$. This makes the pressure
applied by the gas weaker since, for $m>0$, the factor $\dot V$ (see
Eq.~\ref{pgas0}) is multiplied by a factor $\dot\varepsilon/V$ (see
Eq.~\ref{pgas}).

%
%
\section{Cluster energy  dissipation in the gas-cluster collision}
\label{apb}

\subsection{$m>0$}
In order to show that the cluster's kinetic energy is not affected
by the gas-cluster collision on the order of approximation we have
chosen, let's consider the sum of the velocities after the first
passage of the gas particle all the way to the inelastic wall: \be
\sum_{i=0}^{N-1}v_{i}''=\left[1-(1+m)\,
\frac{A}{2}\left|\frac{V}{g_0}\right|^{m}\right]
\sum_{i=1}^{N}v_{i}-\,\frac{A}{2}\left|\frac{V}{g_0}\right|^{m}NV.
\label{p01} \ee Our calculations are carried out to order
$N|V/g_0|^m$ in the development of the coefficient of restitution,
in the spirit of Eq.~\ref{init}. We suppressed terms coming from
orders smaller than $N|V/g_0|^m$.

Let's also consider the sum of the velocities after the passage
back of the gas particle \be \sum_{i=1}^{N}v_{i}''''=\left[1-(1+m)\,
\frac{A}{2}\left|\frac{V}{g_0}\right|^{m}\right]
\sum_{i=0}^{N-1}v_{i}''-\,\frac{A}{2}\left|\frac{V}{g_0}\right|^{m}NV.
\label{p02}
\ee

These equations can be added up giving
\be
\sum_{i=1}^{N}v_{i}''''-\sum_{i=1}^{N}v_{i}=
-(1+m)\,A\,\left|\frac{V}{g_0}\right|^{m}\sum_{i=1}^{N}v_{i},
\label{p03}
\ee
yielding the pressure exerted on the cluster by the gas.

We can proceed along similar lines for the sum of square velocities
and obtain \be \sum_{i=0}^{N-1}(v_{i}'')^{2}=\sum_{i=1}^{N}v_{i}^{2}
-AV\left|\frac{V}{g_0}\right|^{m}\sum_{i=1}^{N}v_{i},
\label{en01}
\ee
and
\begin{eqnarray}
\sum_{i=1}^{N}(v_{i}'''')^{2} &=& \sum_{i=0}^{N-1}(v_{i}'')^{2}
        +AV\left|\frac{V}{g_0}\right|^{m}\sum_{i=0}^{N-1}v_{i}''.
\label{en02}
\end{eqnarray}

Equations~\ref{p01},~\ref{en01} and~\ref{en02} show that the kinetic
energy of the cluster remains the same: \be
\sum_{i=1}^{N}(v_{i}'''')^{2}=\sum_{i=1}^{N}v_{i}^{2} + {\cal
O}(\sigma^2|V/g_0|^{2m}).  \ee

\subsection{$m=0$}
As shown in appendix~\ref{apa2}, the gas grain pumps momentum into the
cluster. We will assume that $NA\ll 1$. This is not too restrictive to our
argument since we will show that a long-time granular
collapse happens for the quasi-elastic velocity-independent
coefficient of restitution case. Thus, it will certainly happen for the case
when $NA$ is large too.

We noticed that as the fast grain collides with the cluster it gives
energy to it by changing the particles' velocities by an amount
proportional to $NA\dot V$. After squaring all cluster particles'
velocities (relative to the center of mass of the cluster), adding
them all up and subtracting the initial value of it, we obtain a rate
of energy, pumped into the cluster, proportional to the product of
$\dot V$ and $\dot\varepsilon$.

\begin{figure}[ht]
\includegraphics[scale = 0.3]{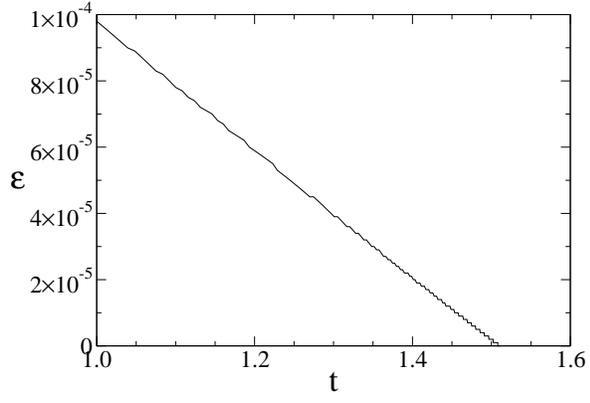}
\caption{Coefficient of restitution independent of the velocities.
The only stable case at all times. The mean-spacing $\varepsilon$ is
measured in units of $L/N$.}
\label{migual0}
\end{figure}

\begin{figure}[ht]
\includegraphics[scale = 0.3]{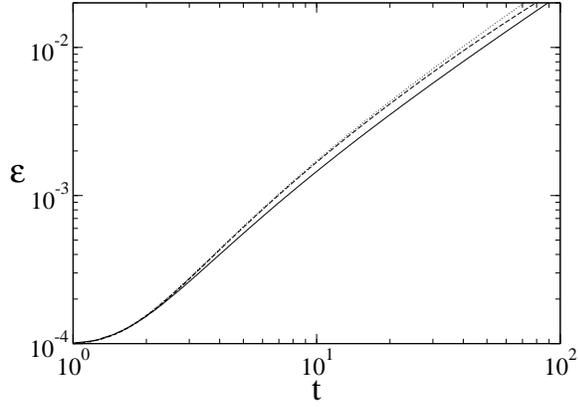}
\caption{Time evolution for a few examples of velocity dependent
coefficients of restitution: full line $m=0.2$ (viscoelastic
model); dashed line $m=1$; dotted line $m=2$. } \label{fig_todas}
\end{figure}

\begin{figure}[ht]
\includegraphics[scale = 0.3, angle = -90]{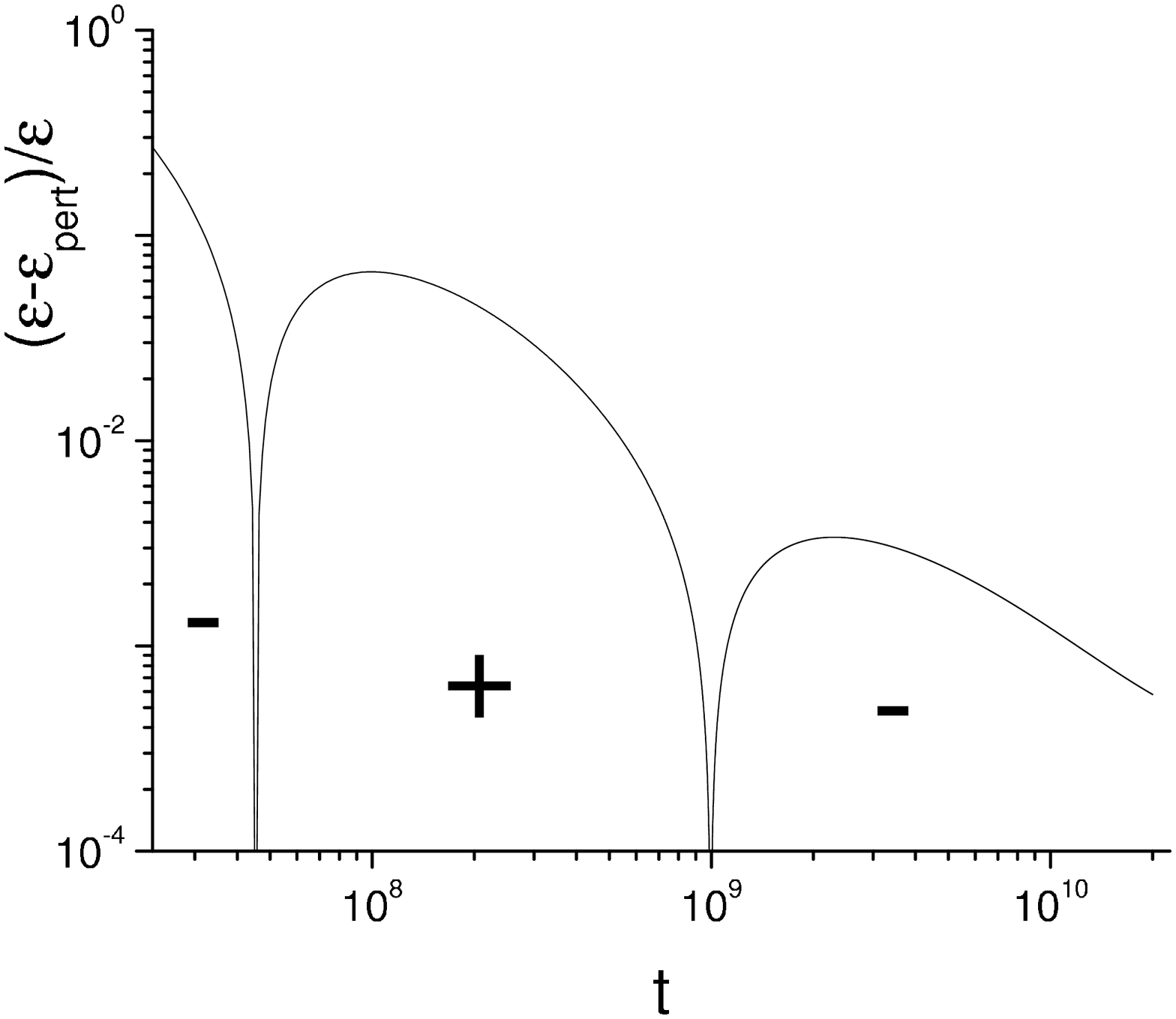}
\caption{Plot of the unperturbed value of $\varepsilon$ minus the
perturbed one, as a function of time, normalized by $\varepsilon$
itself. It is consistent with Eq.~\ref{HO}, a
damped-harmonic-oscillator-like equation. The sign corresponds to
whether the oscillation has a positive or a negative value, since the
plot corresponds to the logarithm of the absolute value of the
difference between the two freely evolving solutions.}
\label{amplia}
\end{figure}

The rate of change of $\sigma$ has two main contributions: a negative
one from internal collisions; and a positive one from gas-cluster
collision. The second one is negligible and we do not take it into
further account. The reason for it goes as follows. Since $|\dot V|\gg
|\dot \varepsilon\dot V|$, if we assume $|\dot \varepsilon\dot V| >
\sigma^2/\varepsilon$ then $\sigma$ will decay more slowly when the
energy pumping term is present but the wall pressure term in
Eq.~\ref{e30} will still be much smaller than the gas one. However, we
can check a posteriori that even when the energy pumping term is not
present, the mean inter-spacing falls at a linear rate in a finite time
(collapse, see figure~\ref{migual0}). It yields
\[
\sigma^2/\varepsilon\sim  |\dot V| \gg |\dot V\dot \varepsilon |.
\]

Thus, we only keep the internal collisions dissipation term in the
equation for $\dot\sigma$:
\begin{equation}
 \dot\sigma\approx -\frac{\sigma^2}{\varepsilon}.\label{dotsigma}
\end{equation}


\newpage

\mbox{ }

\newpage

\newpage

\begin{thebibliography}{100}

\bibitem{PT1} H.M. Jaeger, S.R. Nagel, and R.P. Behringer, {\em
Rev. Mod. Phys.}  {\bf 68}, 1259 (1996); T. Shinbrot and F.J. Muzzio,
{\em Nature} {\bf 410}, 251 (2001); Phys. Today {\bf 53}, 25 (2000);
H.J. Herrmann, {\em Physica A} {\bf 313}, 188 (2002).

\bibitem{duran} J. Duran, {\em Sands, Powders and Grains - An Introduction
to the Physics of Granular Materials}, Springer (1999).

\bibitem{enge1} B.J. Ennis, J. Green, and R. Davies,Chemical
Engineering Progress {\bf 90}, 32 (1994).

\bibitem{mobius} M.E. Mobius, B.E. Lauderdale, S.R. Nagel and
H.M. Jaeger, {\em Nature} {\bf 414}, 6861 (2001); M.E. Mobius,
X. Cheng, G.S. Karczmar, S.R. Nagel and H.M. Jaeger, {\em
Phys. Rev. Lett.} {\bf 93}, 198001 (2004).

\bibitem{feitosa} K. Feitosa and N. Menon, {\em Phys. Rev. Lett.} {\bf
88}, 198301 (2002).

\bibitem{breyT}  J.J. Brey, J.W. Dufty and A. Santos, {\em
J. Stat. Phys.} {\bf 97}, 281 (1997); P.A. Martin, J. Piasecki, {\em
Europhys. Lett.}, 613 (1999).


\bibitem{brey1} V. Garz\'o and J. Dufty, {\em Phys. Rev. E} {\bf 60}
(1999) 5706.


\bibitem{GHZ} I. Goldhirsch and G. Zanetti, {\em Phys. Rev. Lett.}
{\bf 70}, 1619 (1993).

\bibitem{tsimring}I.S. Aranson and L.S. Tsimring, cond-mat/0507419.


\bibitem{ben-naim} E. Ben-Naim, J. B. Knight, E. R. Nowak,
H. M. Jaeger and S. R. Naegel, {\em Physica D} {\bf 123}, 380 (1998).

\bibitem{nie1} X. Nie, E. Ben-Naim and S.Y. Chen, {\em
Phys. Rev. Lett.}  {\bf 89}, 204301 (2002).

\bibitem{haff} P.K. Haff, {\em J. Fluid Mech.} {\bf 134} (1983) 401.

\bibitem{tan} M.-L. Tan and I. Goldhirsch, {\em Phys. Rev. Lett.}
{\bf 81}, 3022 (1998).

\bibitem{precorr} T. P\"oschel, N.V. Brilliantov and T. Schwager,
{\em Int. J. Mod. Phys. C} {\bf 13}, 1263 (2002).

\bibitem{kada2} Y. Du, H. Li, L. P. Kadanoff, {\em Phys. Rev. Lett.}
{\bf 74}, 1268 (1995).

\bibitem{ertas2} T.C. Halsey and D Erta\c{s}, cond-mat/0506170.

\bibitem{ertas1} D. Erta\c{s} and T.C. Halsey, {\em Europhys. Lett.}
{\bf 60}, 931 (2002); L.E. Silbert, D Erta\c{s}, G.S. Grest, T.C. Halsey,
D. Levine and S.J. Plimpton, {\em Phys. Rev. E} {\bf 64}, 051302-1
(2001).

\bibitem{brey4} J. J. Brey, F. Moreno, and J. W. Dufty,
{\em Phys. Rev. E} {\bf 54}, 445 (1998).

\bibitem{viscoelastic} G. Kuwabara and K. Kono, {\em
Jpn. J. Appl. Phys.}  {\bf 26}, 1230 (1987); N.V. Brilliantov,
F. Spahn, J.-M. Hertzsch and T. P\"oschel, {\em Phys. Rev. E} {\bf
53}, 5382 (1986); W.A.M. Morgado and I. Oppenheim, {\em Phys. Rev. E}
{\bf 55}, 1940 (1997).

\bibitem{tpnbsw} T. P\"oschel, N.V. Brilliantov and T. Schwager,
{\em Physica A} {\bf 325}, 274 (2003).

\bibitem{edson} W.A.M. Morgado and E. Vernek, {\em
Int. J. Mod. Phys. B} {\bf 18}, 1 (2004).

\bibitem{kada1} L. P. Kadanoff, {\em Rev. Mod. Phys.} {\bf 71}, 435
(1999).

\bibitem{cordero} J.M. Pasini and P. Cordero, {\em Phys. Rev. E} {\bf
63}, 041302 (2001).

\bibitem{sela} N. Sela, and I. Goldhirsch, {\em Phys. Fluids}
{\bf 7}, 507 (1995).

\bibitem{chap} S. Chapman and T.G. Cowling, Mathematical Theory of
Non-Uniform Gases, The; Third Edition, Cambridge University Press,
1995.

\bibitem{poeschel33} N.V. Brilliantov, F. Spahn, J.-M. Hertzsch and
T. P\"oschel, {\em Phys. Rev. E} {\bf 53}, 5382 (1996);
N.V. Brilliantov and T. P\"oschel, {\em Phys. Rev. E} {\bf 61}, 5573
(2000) ; W.A.M. Morgado and I. Oppenheim, {\em Physica A} {\bf 246},
547 (1997).

\bibitem{footnote1} From Eqs.~\ref{e1} and~\ref{e2}, we can write
$\frac{dV^{-1-m}}{dt}=\frac{1+m}{L}$ and $\frac{d\sigma^{-1-m}}{dt}
=\frac{1+m}{\varepsilon} =\frac{1+m}{L}\frac{L}{\varepsilon}\gg
\frac{1+m}{L}$. Hence, $\sigma^{-1-m}$ grows much faster than
$V^{-1-m}$, so $\sigma$ goes to zero faster than $V$.

\bibitem{doolen} E. Ben-Naim, S.Y. Chen, G.D. Doolen and S. Redner,
{\em Phys. Rev. Lett.} {\bf 83}, 4069 (1999).

\end{thebibliography}
\end{document}